\documentclass[prb,cha,onecolumn,superscriptaddress,preprintnumbers,showpacs, amsmath,amssymb]{revtex4-1}
\usepackage{bm}
\usepackage[colorlinks=true,linkcolor=blue,citecolor=blue]{hyperref}
\usepackage{amsmath}
\usepackage{amssymb}
\usepackage{amsthm}
\usepackage{amsfonts}
\usepackage{enumerate}
\usepackage{latexsym}
\usepackage{ifpdf}
\usepackage{graphicx}
\usepackage{makeidx}

\usepackage{scalerel,amssymb,graphicx}

\def\msquare{\mathord{\scalebox{0.4}[0.4]{\scalerel*{\Box}{\strut}}}}

\expandafter\ifx\csname package@font\endcsname\relax\else
 \expandafter\expandafter
 \expandafter\usepackage
 \expandafter\expandafter
 \expandafter{\csname package@font\endcsname}
 \fi
\usepackage{color}
\usepackage{times}

\begin{document}

\title{Evolution of ferromagnetism in two-dimensional electron gas of LaTiO$_3$/SrTiO$_3$}

\author{Fangdi Wen}
\email{fw113@physics.rutgers.edu}
\affiliation{Department of Physics and Astronomy, Rutgers University, Piscataway, NJ 08854, USA}
\author{Yanwei Cao}
\email{ywcao@nimte.ac.cn}
\affiliation{Department of Physics and Astronomy, Rutgers University, Piscataway, NJ 08854, USA}
\affiliation{Ningbo Institute of Materials Technology and Engineering, Chinese Academy of Sciences, Ningbo,
Zhejiang 315201, China}
\author{Xiaoran Liu}
\affiliation{Department of Physics and Astronomy, Rutgers University, Piscataway, NJ 08854, USA}
\author{B. Pal}
\affiliation{Department of Physics and Astronomy, Rutgers University, Piscataway, NJ 08854, USA}
\author{S. Middey}
\affiliation{Department of Physics, Indian Institute of Science, Bangalore 560012, India}
\author{M. Kareev}
\affiliation{Department of Physics and Astronomy, Rutgers University, Piscataway, NJ 08854, USA}
\author{J. Chakhalian}
\affiliation{Department of Physics and Astronomy, Rutgers University, Piscataway, NJ 08854, USA}

\date{\today}

\begin{abstract}
Understanding, creating, and manipulating spin polarization of two-dimensional electron gases at complex oxide interfaces presents an experimental challenge. For example, despite  almost a decade long research effort, the microscopic origin of  ferromagnetism in LaAlO$_3$/SrTiO$_3$ heterojunction is still an open question. Here, by using  a prototypical two-dimensional electron gas (2DEG)\ which  emerges at the interface between  band insulator SrTiO$_3$ and antiferromagnetic Mott insulator LaTiO$_3$, the experiment reveals the evidence for magnetic phase separation in   hole-doped Ti $d^{1}$ t$_{2g}$ system resulting in   spin-polarized 2DEG. 
 The details of electronic and magnetic properties of the 2DEG were investigated by temperature-dependent d.c. transport, angle-dependent X-ray photoemission spectroscopy, and temperature-dependent magnetoresistance. The observation of clear  hysteresis in magnetotransport at low magnetic fields implies  spin-polarization  from  magnetic islands in the hole rich LaTiO$_3$  near the interface. These findings emphasize the role of  magnetic instabilities in doped Mott insulators  thus providing another  path for designing all-oxide   structures relevant to  spintronics applications.          
\end{abstract}

\maketitle

Utilizing the spin and charge degrees of freedom  is a basic idea for spintronics applications, the study of which is fundamentally important for both basic science and device applications. \cite{RMP-2004-IZ,ARCMP-2010-SB,NM-2012-FP,RPP-2015-ME} The past decade has witnessed a sharp rise of interest to complex oxide heterostructures and a great development of deposition techniques at the atomic scale, \cite{RMP-2014-JC,NM-2012-HH,NM-2012-JC,MRS-2013-FMG,ARMR-2014-SS,JPDAP-2016-ML,NJP-2014-LB,MRS-2008-JM,ARMR-DS-2007} resulting  in plethora of  interesting phenomena emerging at complex oxide interfaces. In addition,  those phenomena provide  exciting opportunities for the design of all-oxide  spintronics application  and field-effect devices. \cite{Science-2010-JM,IE-2007-MB,AML-2014-MP,PRL-2012-NR,PRA-2015-CW,RMP-2017-FH} Among these, understanding and control of two-dimensional electron gases (2DEGs) with spin-polarized carriers  have attracted tremendous attention, \cite{PRL-2012-NR,PRB-2017-JB,NM-2017-RO,NC-2016-WL,APL-2011-PM,PRB-2013-CJ,RMP-2017-FH,APL-2013-MK,APL-2014-AS,PRL-2016-YC,APL-2013-JD,PRX-2015-HI,PS-2013-AN} and yet remains a distinct challenge for the experimentalists.

In the past decade, the microscopic origin of ferromagnetism at LaAlO$_3$/SrTiO$_3$ has been actively investigated and yet not  fully understood. \cite{NM-2007-AB,PRL-2011-DD,PRL-2011-MF,NC-2011-A,NC-2012-BK} Specifically, since single crystal SrTiO$_3$ substrates with unavoidable defects and impurities can shows some signature of  ferromagnetism,\cite{JPCM-2016-JC} it is an open question whether the observation of ferromagnetism at the interfaces is an \textit{intrinsic} property of 2DEG, or is more of a result of unavoidable defects/impurities from the sample preparation procedure. \cite{PRL-2011-MF,NC-2012-BK}
Alternatively, ferromagnetic layers have been used to induce the spin-polarization in  2DEGs. For example, ferromagnetic (FM)\ rare-earth titanate GdTiO$_3$ (T$_C$ $\sim$ 32K) was used as a charge/spin donor in GdTiO$_3$/SrTiO$_3$ heterojunctions \cite{APL-2011-PM,PRB-2013-CJ}, magnetic Co layer  was shown to  act as a spin injection source in multilayer heterostructure Co/LaAlO$_3$/SrTiO$_3$, and FM\ YTiO3 layer was explicitly  introduced in to   3-color YTiO$_3$/SrTiO$_3$/LaTiO$_3$ hetrostructure  . \cite{APL-2014-AS,PRX-2015-HI} Following this  route,  one can naturally ask  whether antiferromagnetic materials (e.g., Mott insulators LaTiO$_3$, NdTiO$_3$, and SmTiO$_3$) can induce spin-polarized 2DEGs in SrTiO$_3$-based interfaces. However, recent  works delivered  a set of mixed answers, e.g. showing the   absence of ferromagnetism in SmTiO$_3$/SrTiO$_3$ heterostructure, \cite{NC-2014-CJ} and demonstrating the presence of ferromagnetism  at the NdTiO$_3$/SrTiO$_3$ interface.  \cite{arXiv-2017-YA} Moreover, no clear signature of ferromagnetism  has been reported for the prototypical  2DEG  of LaTiO$_3$/SrTiO$_3$. \cite{Nature-2002-AO}

To  investigate this  issue, we synthesized a series of high-quality SrTiO$_3$/LaTiO$_3$ (STO/LTO) heterostructures with variable thickness of the  LaTiO$_3$ layer by pulsed laser deposition (PLD). The crystal structure was confirmed by X-ray diffraction (XRD) and the electronic and magnetic properties of 2DEG were investigated by temperature-dependent d.c. transport, Hall measurement, angle-dependent X-ray photoemission spectroscopy (XPS), and temperature-dependent magnetoresistance (MR). The combination of probes revealed  clear hysteretic behavior of MR is only observed at the interfaces of STO/LTO\ with thicker LTO layer (20 u.c.) and absent for the   thinner LaTiO$_3$ layer (3 u.c.). Based on the observation we conjecture how the  FM behavior of 2DEG can be resulted from the spin-flip scattering between conduction electrons and  magnetic islands with localized electrons, with an additional contribution to canted ferromagnetism from the inner  antiferromagnetic  atomic planes of LTO layer. Our results show alternative way to control the magnetism of 2DEGs in titanate-based interfaces and should be important for designing all-oxide-based spintronics.

\begin{figure*}[h]
\includegraphics[width=0.8\textwidth]{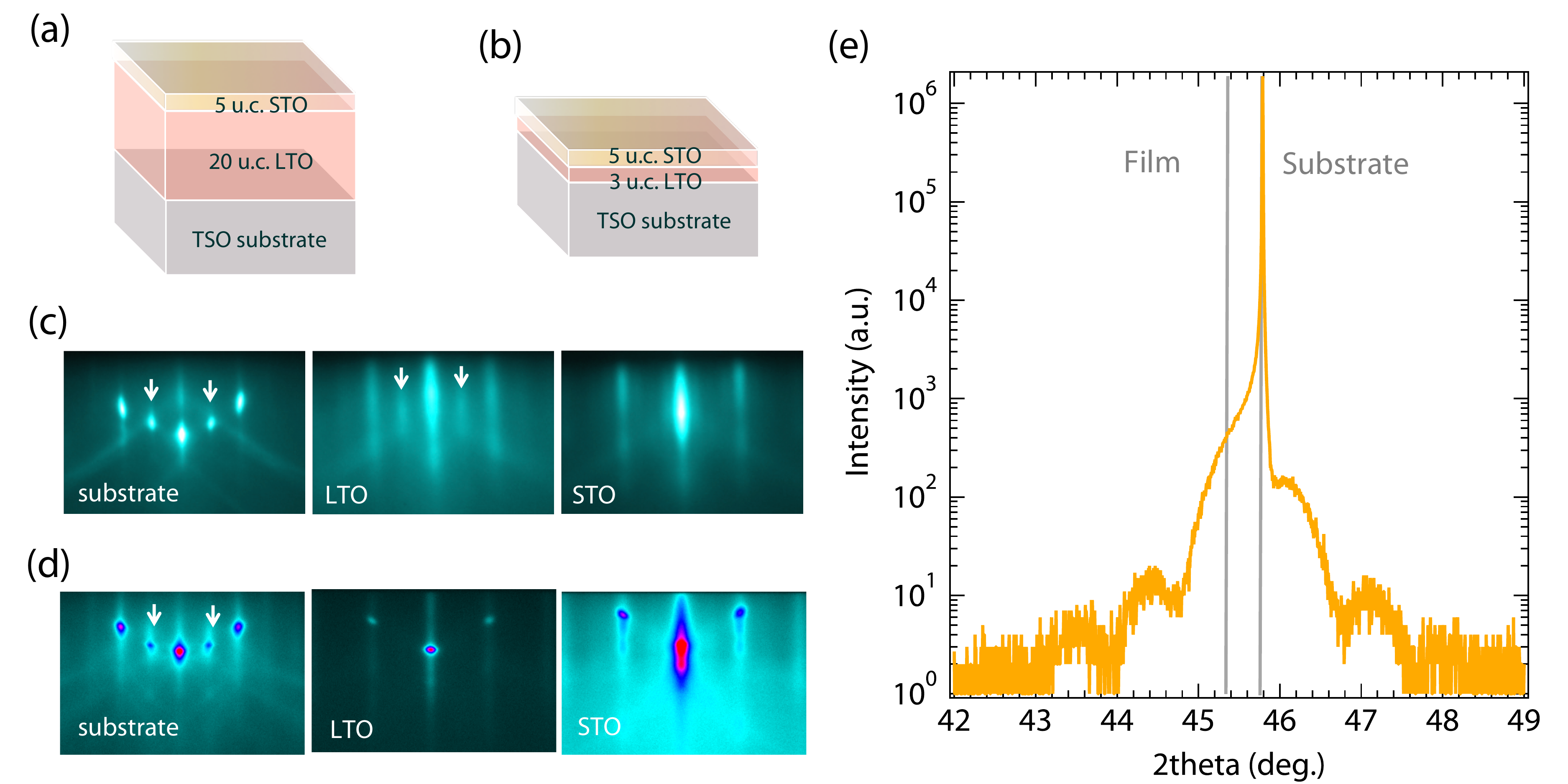}
\caption{\label{} (a) and (b) Sketch of 5u.c.SrTiO$_3$/nu.c.LaTiO$_3$ (n = 20 and 3, u.c. = unit cells) heterostructures on TbScO$_3$ substrates. (c) RHEED patterns of 5STO/20LTO during growth for TbScO$_3$ substrate, LaTiO$_3$, and SrTiO$_3$, respectively. Yellow arrows indicate the half-order-peaks. (d) RHEED patterns of 5STO/3LTO during growth. (e) XRD of 5STO/20LTO sample. The intensity is plotted on a log scale, and the film peak is to the left of substrate peak.  }
\end{figure*}

As shown in  inset of Fig. 1. (a) and (b), in all the designed samples, STO  is a top layer and  grown on the thicker LTO layer resulting in the  2DEG formed at the STO/LTO interface. Specifically, high-quality 5u.c. SrTiO$_3$/n u.c. LaTiO$_3$ (n = 3, 20, 5STO/nLTO after, and u.c. = unit cells) heterostructures on TbScO$_3$ (110)-oriented (orthorhombic notation) single crystal substrates (5 $\times$ 5 $\times$ 0.5 mm$^3$) were epitaxially synthesized by PLD using a KrF excimer laser operating at $\lambda$ = 248 nm and 2 Hz pulse rate. These stated growth condition can be found in our previous reports.\cite{APL-2013-MK,PRL-2016-YC} Figures 1 (c)-(d) exhibit typical \textit{in-situ} RHEED patterns with distinct specular and off-specular  reflections  distributed on the Laue rings which attest for  the high crystallinity and flatness of the heterostructures. Notice,  the growth sequence of 5STO/$n$LTO samples is quite different from the previously reported results in which LTO is always the top layer  grown on STO substrates. \cite{PRB-2012-AR,NC-2010-JB,PRB-2010-FW} The main reason for the inverted layer sequence (i.e. STO/LTO vs. LTO/STO)\ is two-fold. First, the  surface of  antiferromagnetic LTO can be easily oxidized and becomes paramagnetic LaTiO$_{3+\delta}$; \cite{PRB-1999-YT} In addition, as previously  reported the STO substrate  itself can host metallicity and magnetism.\cite{PRL-2014-XL,JPCM-2016-JC} To avoid this,  STO is replaces by  inert  TbScO$_3$ in our samples.   Structural properties of the film were further investigated by high-resolution X-ray diffraction (XRD) in the vicinity of TSO (002) reflection. As shown in Fig. 1.(e), left to the sharp peak of the substrate, the broad (002) reflection of the film is clearly observed with well-defined Kiessig fringes. From the angular interval of the fringes, the total thickness of the film is determined $\sim$ 9.31 nm. This result agrees well with the estimated value according to the numbers of intensity oscillations of RHEED during the deposition, corroborating the high quality of the film.

\begin{figure*}[]
\includegraphics[width=0.8\textwidth]{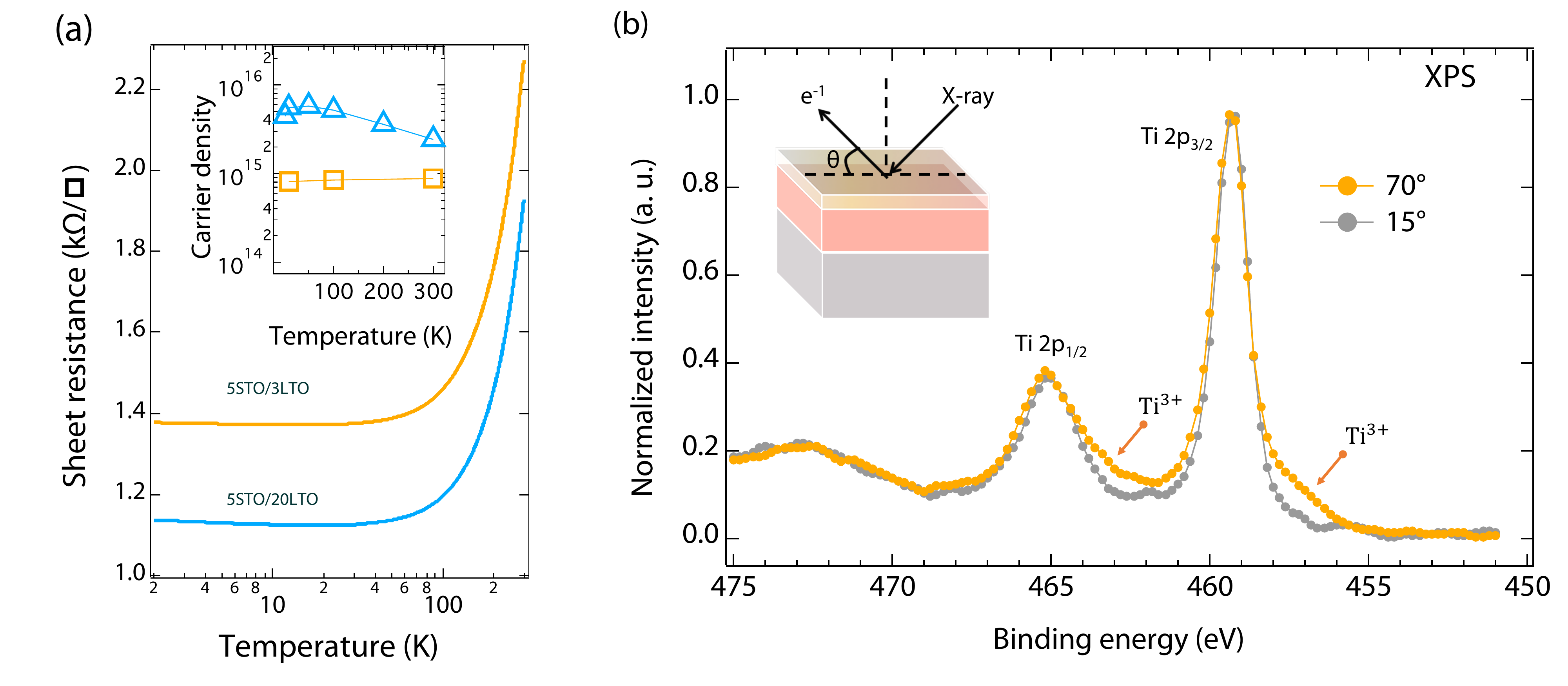}
\caption{ \label{}  (a) Temperature-dependent sheet resistances of 5STO/20LTO and 5STO/3LTO. (b) Angle-dependent Ti $2p$ core level XPS on 5STO/20LTO at room temperature. $\theta$ is the angle between sample surface and the axis of acceptance angle of electron energy analyzer.}
\end{figure*}

With the availability of high  quality  samples, we have  investigated  in detail the properties of 2DEG in 5STO/$n$LTO ($n$ = 3 and  20). First, to confirm the formation of a 2DEG, temperature-dependent resistance measurements were carried out. As shown in Fig. 2~(a), the sheet resistance of the  5STO/20LTO sample decreases from $\sim$ 1.9 k$\Omega/\msquare$ at 300 K to $\sim$ 1.1 k$\Omega/\msquare$ at 2 K , indicating a typically metallic behavior. Similarly, as expected, the temperature dependent resistivity of 5STO/3LTO also shows metallic behavior in the same \textit{T}-range. The existence of 2DEG was proved by measuring Hall effect on both 5STO/20LTO and 5STO/3LTO sample in different temperature. By plotting the temperature dependence of carrier density, we got a carrier density close to  $4\times 10^{15} cm^{-2}$  for 5STO/20LTO and $8\times 10^{14} cm^{-2}$  for 5STO/3LTO stable under changing temperature. This data is close to the estimated carrier density $3\times10^{14} cm^{-2}$ for half electron doping. This number does not change a lot when the temperature changes, which is also a typical property of 2DEG, as shown in the inset of Fig. 2. (a).
To further investigate the electronic structure near the interface of 5STO/20LTO, we performed angle-dependent XPS measurements. Fig. 2. (b) shows typical Ti $2p$ core level spectra acquired at two acceptance angles of the analyzer. Due to the small value  of  the  inelastic mean free path ($\sim1.5$ nm for 1 keV), \cite{PRL-2014-BD} the probing depth is extremely close to the surface for small angle ($\theta$ = 15$^{\circ}$), while for the larger angle ($\theta$ = 75$^{\circ}$), it is more sensitive to deeper parts of the structure. As seen in Fig. 2. (b), the  main two features near 459.3 eV and 465.2 eV can be  assigned to Ti$^{4+}$ $2p_{3/2}$ and $2p_{1/2}$ peaks, respectively. \cite{APL-2011-PM} Most importantly, in contrast to  the small-angle spectra  ($\theta$ = 15$^{\circ}$, surface sensitive),  the spectra acquired at  $\theta$ = 70$^{\circ}$  exhibit two pronounced shoulders near 457 eV and 463 eV (marked by  arrows in Fig. 2.(b)).  Those energy  positions corresponds to  the contribution of Ti$^{3+}$ $2p_{3/2}$ and $2p_{1/2}$ peaks. \cite{APL-2011-PM} The relative intensity of Ti$^{3+}$ increases as we go from lower to higher angle. It suggests that the surface is more Ti$^{4+}$ rich and the relative concentration of Ti$^{3+}$ increases as with go deep inside the film. These results are consistent with our sample design as 5 u.c. of SrTiO$_3$ is deposited on top of 20 u.c. LaTiO$_3$ in this bi-layer system.

\begin{figure*}[]
\includegraphics[width=0.5\textwidth]{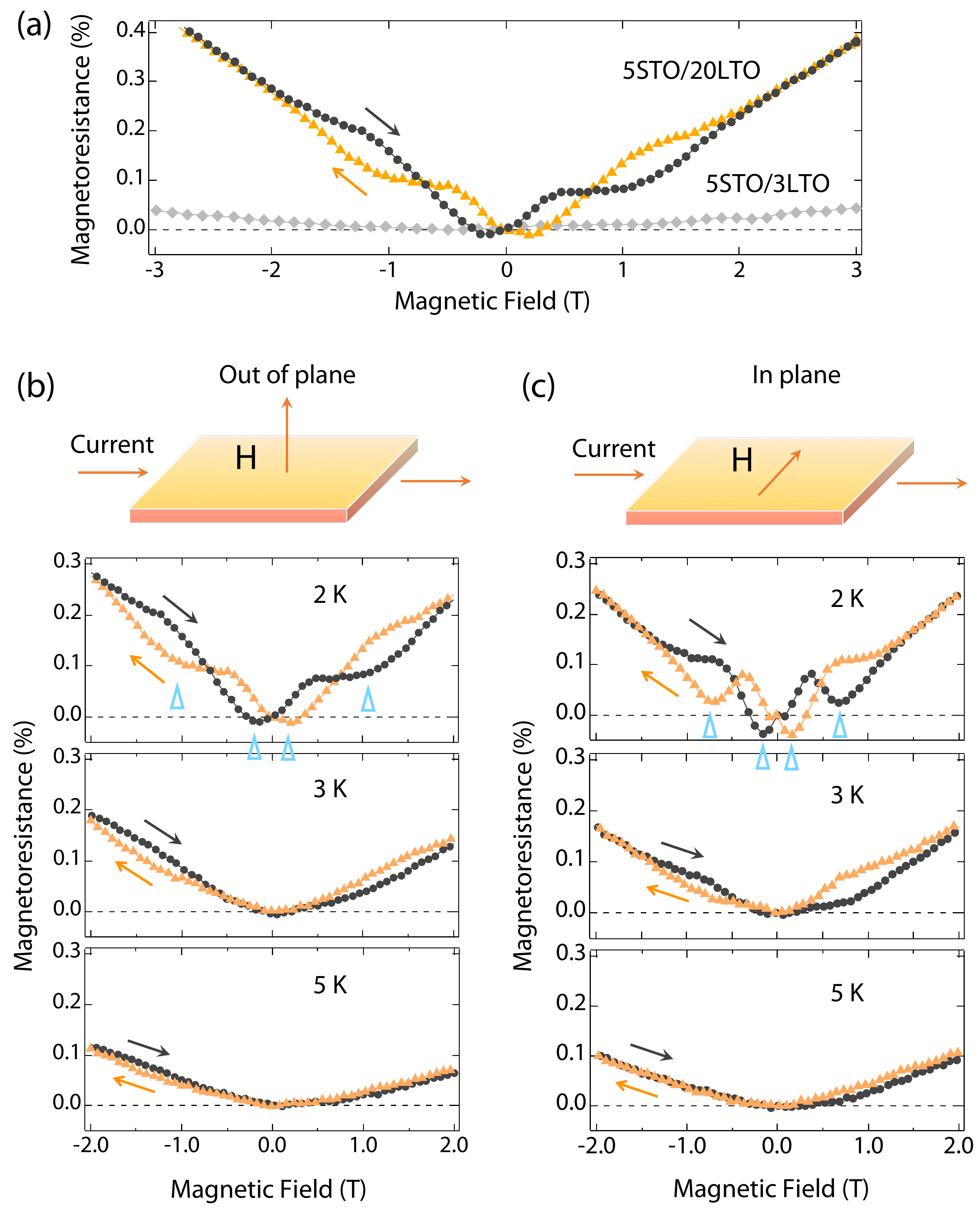}
\caption{ \label{} (a) Comparison of magnetoresistance between 5STO/20LTO and 5STO/3LTO heterostructures at temperature 2 K with out-of-plane magnetic field. (b) and (c) Temperature-dependent magnetoresistance of 5STO/20LTO with out-of-plane and in-plane magnetic fields, respectively. Black and orange arrows indicate the sweep directions of the external magnetic field, whereas blue triangles mark the critical magnetic field of magnetoresistance. Here, the magnitude of magnetoresistance (\textit{MR}) is defined as \textit{MR} = [\textit{R(H)} - \textit{R(0)}] / \textit{R(0)} $\times$ 100 $\%$, where \textit{R(H)} represents the sheet resistance under external field \textit{H}.}
\end{figure*}

Next, we explore the magnetic properties of 5STO/$n$LTO. Figure 3 shows the  magnetoresistance (\textit{MR}) data for 5STO/20LTO taken from 2K to 5 K with an applied external magnetic field along in-plane and out-of-plane directions, respectively. As seen in Fig. 3(a), in sharp  contrast to  the  conventional positive magneto-resistance in 5STO/3LTO, a clear hysteresis was found in 5STO/20LTO with out-of-plane magnetic field at 2 K. This emergent feature is  surprising considering that the thickness is the only changed factor. In order to gain further  insight into  this phenomenon, we measured temperature-dependent magneto-resistance on 5STO/20LTO with both out-of-plane (Fig.~3(b)) and in-plane (Fig.~3(c)) magnetic fields. As shown in Fig.~3(c), besides the usual  positive magneto-resistance  above 1.5 T, the dominant feature here is the presence of hysteresis  under the moderate magnetic fields, implying  the formation of ferromagnetism in 2DEG. \cite{NM-2007-AB,PRL-2011-DD,arXiv-2017-YA} Interestingly, unlike a simple line shape of hysteresis  reported for SrTiO$_3$/NdTiO$_3$, \cite{arXiv-2017-YA} our heterostructure shows a more complex line shape with  two critical magnetic fields of $\sim$ $\pm$ 0.2 and $\pm$ 0.7 T  (marked by blue triangles in Fig.~3(c) at 2 K).  Upon increasing the temperature from 2 to 3 K, the hysteresis becomes strongly suppressed but the second critical field of $\sim$ $\pm$ 0.7 T still survives. This feature disappears completely  once the temperature exceeds 5 K. The observed MR\ behavior  under out-of-plane magnetic field [Fig.~3(b)] is analogous to  that of  in-plane magnetic field [Fig.~3(c)].  

\begin{figure*}[]
\includegraphics[width=0.5\textwidth]{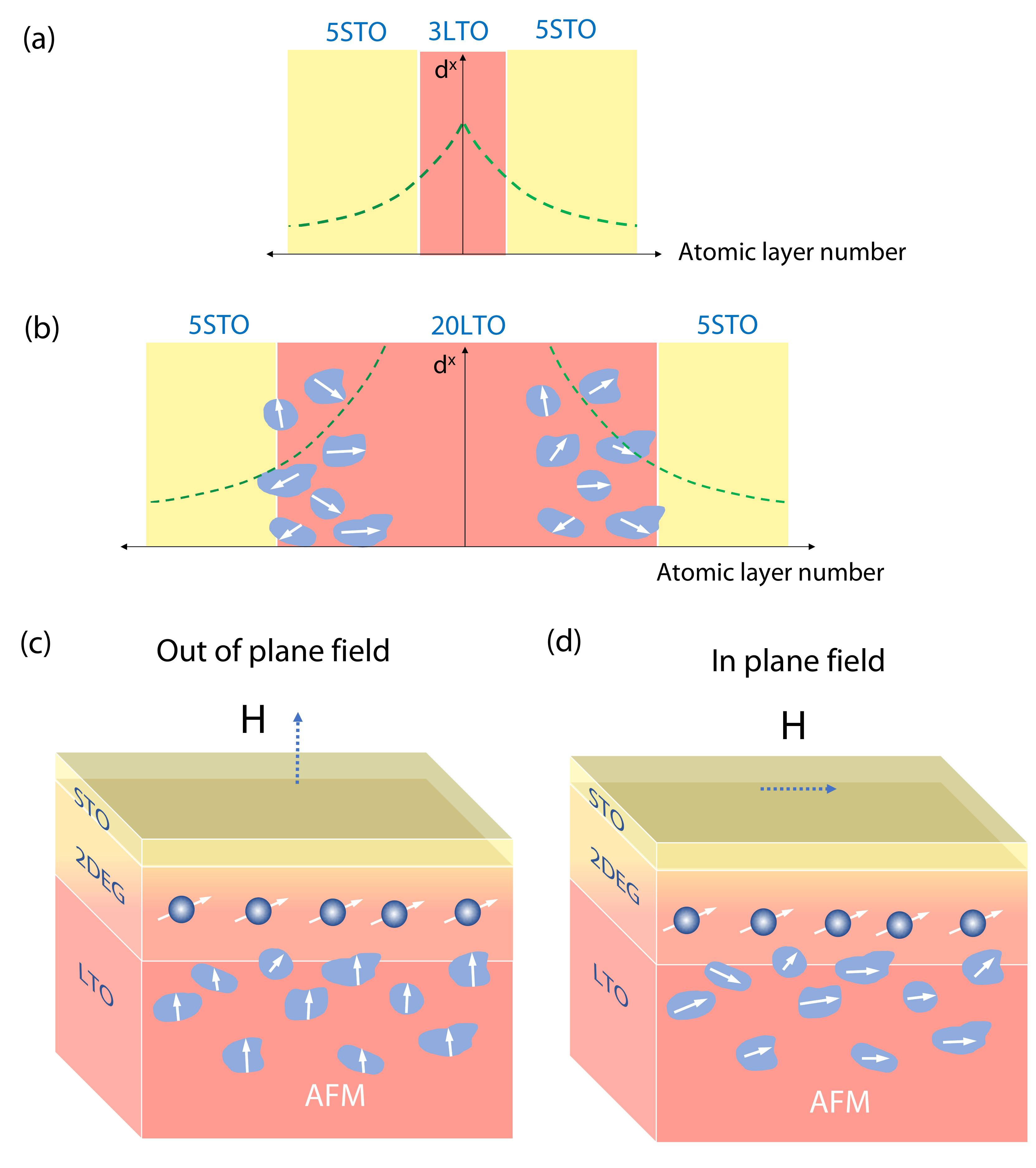}
\caption{ \label{} (a) Schematic of charge decay of 5STO/3LTO. (b) Schematic of charge decay of 5STO/20LTO with the formation of magnetic islands in LaTiO$_3$ layers. (c) and (d) Schematic of origin of spin-polarized 2DEG (blue balls) in 5STO/20LTO interface with out-of-plane and in-plane magnetic fields, respectively. Blue domains show magnetic islands with random moment near the interface, whereas the white arrows mark the directions of magnetic moments.}
\end{figure*}

In general, MR can be attributed to various sources including the Lorentz force, spin scattering, anisotropicity, weak localization, and so on. \cite{JAP-2006-PS} As well known, with the application of an external magnetic field, the Lorentz force can reduce the mobility of conduction carriers and increase the amplitude of sheet resistance, resulting in a classical positive and non-saturating magneto-resistance; In this mechanism, \textit{MR} $\sim H^2$. \cite{EPL-2008-GG} As clearly  seen in Fig. 3, the effect of Lorentz force is distinct for the  field larger than 2 T, giving rise to the parabolic \textit{MR} contribution. On the other hand, under low magnetic field the contribution of spin scattering becomes significant leading to the hysteretic behavior of \textit{MR}.\cite{NM-2007-AB} From  the similarity between the line shape of \textit{MR} under  in- and out-of-plane magnetic field [Fig.~3(b) and 3(c)] we can rule out the leading effect of anisotropy for observed hysteresis.

Other than general factors, we discussed a possible microscopic origin of spin polarization (or ferromagnetism)\ in the 2DEG in 5STO/20LTO. The observed hysteric \textit{MR}\  may  arise from two contributions: (1) the spin-flip scattering between 2DEG and  magnetic islands formed near the interface and (2) the magnetic interaction of 2DEG with canted ferromagnetism from LTO layers. 
Theoretically, it has been predicted that in a hole doped  $3d^{1}$ t$_{2g}$ Mott insulator (e.g. LaTiO$_3$ or  YTiO$_3$) away from half-filling, the system becomes thermodynamically unstable which results in a phase separation into antiferromagnetic, ferromagnetic or paramagnetic Fermi liquid regions depending on the doping level. \cite{CUP-2010-DK,PRX-2015-CY} 
Furthermore, it has been proposed that  in the case of SrTiO$_3$/NdTiO$_3$ the  phase-separation mechanism may   transfer  such spin-polarization into 2DEG. \cite{arXiv-2017-YA} 
Similar to the conclusion people got in the SrTiO$_3$/NdTiO$_3$ system, as schematically shown in Fig.~4, due to interfacial charge transfer from LTO into STO layers, the Ti 3$d$ band is less than half filled in LaTi$^{(3+/4+)}$O$_3$. This less than half filled band structure in turn leads to the phase separation in hole-doped Mott insulator LaTi$^{(3+/4+)}$O$_3$ with ferro- or ferrimagnetic islands formed  near the STO/LTO interface. \cite{CUP-2010-DK} 
However,  when the thickness of the LTO layer approaches its ultra thin limit, the doping level was effectively increased, and the hole-doped LTO will favor paramagnetic Fermi liquid. 
This explains the absence of hysteresis in 5STO/3LTO (see Fig. 4 (a)).
On the other hand, for  5STO/20LTO whose LTO layer is significantly thicker  (20 u. c. vs. 3 u. c. ), the spin-flip scattering between conduction electrons and magnetic islands can mediate the spin-polarization in 2DEG, (see Fig.s 4(b)-(d)). \cite{arXiv-2017-YA,nature_comm,spin_flip_SC}
 As shown in Fig. 4. (c), after application of the external magnetic field, the orientation of magnetic moments in the magnetic islands is expected to follow the direction of the magnetic field in both in-plane and out-of-plane case, which is consistent with the observation of isotropic critical field of $\sim$ $\pm$ 0.2 T for both  in- and out-of-plane magnetic fields. In addition to  the unconventional  \textit{MR}, the other interesting  feature  is the  prepense of  extra minima.  The canting of the AFM ordering affect the distribution and orientation of the magnetic islands that interact with the mobile electrons at the interface, and leads to a peak shift of the additional dip. Specifically, when we applied out-of-plane field, the different dip location ($\sim$ 1T for out of plane whereas $\sim$ 0.7T for in-plane) indicates the anisotropic axis of the spins in LTO tends to align more along the out-of-plane direction. \cite{nature_comm} 

In summary, we report on the evolution of ferromagnetism in the  2DEG formed at the  interface of   5STO/\textit{n}LTO (n=20 and 3) heterostructures. The RHEED and XRD were present to confirm our crystal structure, and temperature-dependent electrical transport, Hall effect and angle-dependent XPS were performed to characterize the mixed valency  of Ti$^{3+/4+}$ near the interface of STO/LTO. By measuring temperature-dependent magnetoresistance, we revealed the presence of  clear hysteresis of magneto-resistance, implying the spin polarization of 2DEG in 5STO/20LTO. This ferromagnetic behavior of the 2DEG  can be attributed primarily to the spin-flip scattering between conduction carriers,  ferro- or ferri- magnetic islands and canted antiferromagnetism near the interface. Our results emphasize the importance and phase instability  for the  correlated oxide interfaces  which is  directly  relevant for designing all-oxide spintronics applications based on doped Mott materials.

\textbf{Acknowledgements}

F. W. was supported by the Claud Lovelace Graduate Fellowship. Y. C. acknowledged the Gordon and Betty Moore Foundations' EPIQS Initiative through ICAM-I2CAM, Grant GBMF5305. X. L. was supported by the Department of Energy Grant No. DE-SC0012375. J. C. acknowledged the support by the Gordon and Betty Moore Foundation EPiQS Initiative through Grant No. GBMF4534. 



\end{document}